\documentclass[preprint2]{aastex}
\usepackage{emulateapj5}
\usepackage{onecolfloat5}
\slugcomment{Submitted to the Astrophysical Journal, 22 January 2002; 
        accepted 21 February 2002.}

\newcommand{\eg}{e.g.}

\newcommand{\<}{$<$}
\newcommand{\etal}{{\it et~al.\/}\ }
\newcommand{\minusone}{$^{-1}$} 
\newcommand{\LOH}{$L_{OH}$}
\newcommand{\LFIR}{$L_{FIR}$}
\newcommand{\I}{\protect\small I \normalsize $\!\!$}
\newcommand{\II}{\protect\small II \normalsize $\!\!$}
\newcommand{\HI}{\mbox{\rm H\I}\ }

\newcommand{\HII}{\mbox{\rm H\II}\ }

\shorttitle{OH Megamaser Luminosity Function}
\shortauthors{Darling \& Giovanelli}

\begin{document}
\twocolumn[
\title{The OH Megamaser Luminosity Function} 
\author{Jeremy Darling \& Riccardo Giovanelli}
\affil{Department of Astronomy and National Astronomy and Ionosphere Center, 
Cornell University, Ithaca,  NY  14853;
darling@astro.cornell.edu; riccardo@astro.cornell.edu}

\begin{abstract}
We present the 1667 MHz OH megamaser luminosity function derived
from a single flux-limited survey.  The Arecibo 
Observatory OH megamaser (OHM) survey has doubled the 
number of
known OH megamasers, and we list the complete catalog of OHMs detected by the 
survey here, including three redetections of known OHMs.  OHMs
are produced in major galaxy mergers which are (ultra)luminous in the 
far-infrared.  The 
OH luminosity function follows a power law in integrated 
line luminosity,  $\Phi \propto L_{OH}^{-0.64}$ Mpc$^{-3}$ dex\minusone,
and is well-sampled for $10^{2.2}L_\odot < L_{OH} < 10^{3.8}L_\odot$.  
The OH luminosity function
is incorporated into predictions of the detectability and areal density 
of OHMs in high redshift OH surveys for 
a variety of current and planned telescopes
and merging evolution
scenarios parameterized by $(1+z)^m$ in the merger rate ranging from
$m=0$ (no evolution) to $m=8$ (extreme evolution).  Up to dozens of
OHMs may be detected per square degree per 50 MHz by a survey reaching
an {\it rms} noise of 100 $\mu$Jy per 0.1 MHz channel.  An adequately sensitive
``OH Deep Field'' would significantly constrain the evolution exponent $m$
even if no detections are made.  In addition to serving as luminous 
tracers of massive mergers, OHMs may also trace highly obscured nuclear
starburst activity and the formation of binary supermassive black holes.
\end{abstract}
]

\section{Introduction}

OH megamasers are luminous masing lines at 1667 and 1665 MHz 
which are at least a million times more luminous than typical OH 
masers associated with compact \HII regions.  OH megamasers (OHMs)
are produced in (ultra)luminous infrared galaxies ([U]LIRGs), major 
galaxy mergers undergoing extreme bursts of circumnuclear star formation.
OH megamasers are especially promising
as tracers of dust obscured star formation and merging galaxies because
they can potentially be observed at high redshifts with modern radio 
telescopes in reasonable integration times, they {\it favor} regions of 
high dust opacity where ultraviolet, optical, and even near-IR
emission can be extremely difficult to detect, and their detection 
automatically provides an accurate redshift measurement.  
The Arecibo Observatory\footnote{The Arecibo
Observatory is part of the National Astronomy and Ionosphere Center, which 
is operated by Cornell University under a cooperative agreement with the
National Science Foundation.} OH megamaser
survey is a flux-limited survey designed to quantify the relationships
between merging galaxies and the OH megamasers (OHMs) they produce with the
goal of using OHMs as luminous tracers of mergers at high redshifts 
(Darling \& Giovanelli 2000, 2001, 2002; hereafter Papers I, II, III).  
Central to the 
application of OHMs as tracers of merging galaxies at various redshifts 
is a measurement of the low redshift OH luminosity function.  

The Arecibo OH megamaser survey is the first survey with adequate statistics
to construct an OH luminosity function (LF) from a flux-limited sample.  
Baan (1991) computed a OH luminosity function from the 48 OHMs which 
were known at
the time under the assumption that they constituted a flux-limited sample
and that the fraction of LIRGs showing detectable OHMs is constant.  
Baan's OH LF has a Schecter-type
profile which is slowly falling from $L_{OH} = 10^0$--$10^2\ L_\odot$, 
has a knee at roughly $10^{2.5}\ L_\odot$, and a steep falloff out to 
$10^4\ L_\odot$.  We are now in a position to recompute the OH LF from
a single complete survey with well-defined selection criteria extracted
from the PSCz, a flux-limited catalog which also has well-defined 
selection criteria 
(Saunders \etal 2000).  An OH LF will be a useful guide for deep surveys
for OHMs which can be related to the merging history of galaxies, the 
dust-obscured star formation history of the Universe, and the production 
of some portions of the low frequency gravitational wave background.

This paper presents an overview of the Arecibo OH megamaser survey, focusing
on the issues pertinent to constructing an OH luminosity function (OH LF) 
from the
survey results.  Selection methods and the complete catalog of detected 
OHMs are presented in \S \ref{sec:survey}.  Section \ref{sec:OHLF} discusses
the methods used to compute the OH luminosity function, presents
the OH LF, and compares it to previous results for OHMs and ULIRGs.  
The OH LF is then applied to the problem of detecting OHMs at high redshift
in \S \ref{sec:detect_hiz} and some discussion is made of the utility
of OHMs as tracers of galaxy evolution, including merging, dust-obscured 
nuclear starbursts, and the formation of binary supermassive black holes.

Note that Papers I--III assume a cosmology with 
$H_\circ = 75$ km s\minusone\ Mpc\minusone, $q_\circ = 0$, 
and $\Omega_\Lambda = 0$ for ease of comparison with previous (U)LIRG 
surveys such as \citet{kim98}.  
This analysis, however, assumes a more likely cosmology which is flat
but accelerating:  $\Omega_M = 0.3$ and $\Omega_\Lambda = 0.7$.  All of the
OHM data presented here have been converted to this cosmology.

\section{The Arecibo OH Megamaser Survey}\label{sec:survey}

Papers I--III present a survey for OHMs
conducted at the Arecibo Observatory covering one quarter of the sky 
to a depth of roughly one gigaparsec.  The survey doubles the number of 
known OH megamasers and quantifies the relationship between luminous 
infrared galaxies and OH megamasers.  The survey builds the 
foundation required to employ OH megamasers
as tracers of major galaxy mergers, dust-obscured star formation, and
the formation of binary supermassive black holes spanning the epoch of 
galaxy evolution to the present.  
Here we present details of the survey relevant to constructing an OH LF:
the candidate selection criteria and the complete catalog of detected 
OHMs with properties converted to a flat $\Omega_\Lambda =0.7$ cosmology.

\subsection{Candidate Selection\label{sec:selection}}

For the Arecibo OH megamaser survey, candidates were selected from the 
Point Source Catalog redshift survey (PSCz; Saunders \etal 2000), 
supplemented by the NASA/IPAC Extragalactic 
Database.\footnote{The NASA/IPAC Extragalactic Database (NED) is operated 
by the Jet Propulsion 
Laboratory, California Institute of Technology, under contract with the 
National Aeronautics and Space Administration.}
The PSCz catalog is a flux-limited ({\it IRAS} $f_{60\mu m} > 0.6$ Jy)
redshift survey of 15,000 {\it IRAS} galaxies over $84\%$ of the sky
(see Saunders \etal 2000).
We select {\it IRAS} sources which are in the Arecibo sky ($0^\circ <
\delta <
37^\circ$), were detected at 60 $\mu$m, and have $0.1\leq  z \leq 0.45$.  
The lower redshift bound is set to avoid local radio frequency interference
(RFI), while the upper bound is set by the bandpass of the wide L-band receiver
at Arecibo, although an effective upper bound is imposed around $z=0.23$
by the radio frequency interference (RFI) environment, as discussed in 
\S \ref{subsec:OHLF}.  
No constraints are placed on FIR colors or luminosity.  The 
redshift requirement limits the number of candidates in the Arecibo sky 
to 311.  The condition that
candidates have $z>0.1$ automatically selects (U)LIRGs if they are included
in the PSCz.  The strong influence of \LFIR\ on OHM fraction
in LIRGs (see Paper III) is the primary reason for our high detection
rate compared to previous surveys (\eg\ Staveley-Smith \etal 1992; 
Baan, Haschick, \& Henkel 1992).

\begin{deluxetable}{lccccrrrrcc}
\tabletypesize{\scriptsize}
\tablecaption{Arecibo Survey OH Megamasers:  Optical Redshifts and 
		FIR Properties 
	\label{detectFIR}}
\tablewidth{0pt}
\tablehead{
\colhead{{\it IRAS} Name} &  \colhead{$\alpha$} & \colhead{$\delta$} & 
\colhead{$z_\odot$} & 
\colhead{Ref} &
\colhead{$cz_\odot$} & 
\colhead{$cz_{CMB}$} & 
\colhead{$D_L$} &
\colhead{$f_{60\mu m}$} &
\colhead{$f_{100\mu m}$} &
\colhead{$\log L_{FIR}$}  \\
\colhead{FSC} & \colhead{B1950} & \colhead{B1950} &\colhead{} &
\colhead{z} & \colhead{km/s} & \colhead{km/s} & \colhead{$h^{-1}_{75}$Mpc} & 
\colhead{Jy} & \colhead{Jy} & \colhead{$h^{-2}_{75} L_\odot$} \\
\colhead{(1)}&\colhead{(2)}&\colhead{(3)}&\colhead{(4)}&\colhead{(5)}&
\colhead{(6)}&\colhead{(7)}&\colhead{(8)}&\colhead{(9)}&\colhead{(10)}&
\colhead{(11)}
}
\startdata
01562+2528 & 01 56 12.0 & +25 27 59 & 0.1658 & 1 & 49707(505)
& 49441(506) & 739(8) & 0.809(\phn57) & 1.62(24) & 11.90 \\
02524+2046 & 02 52 26.8 & +20 46 54 & 0.1815 & 1 & 54421(125)
& 54213(129) & 788(2) & 0.958(\phn77) & \<\phn4.79 & 11.81--12.28 \\
03521+0028 & 03 52 08.5 & +00 28 21 & 0.1522 & 2 & 45622(138)
& 45501(142) & 674(2) & 2.638(237) & 3.83(34) & 12.28 \\
03566+1647 & 03 56 37.8 & +16 47 57 & 0.1335 & 1 & 40033(\phn54)
& 39911(\phn65) & 585(1) & 0.730(\phn66) & \<\phn2.37 & 11.40--11.75 \\
04121+0223 & 04 12 10.5 & +02 23 12 & 0.1216 & 3 & 36454(250)
& 36362(253) & 529(4) & 0.889(\phn62) & \<\phn2.15 & 11.40--11.69 \\
06487+2208 & 06 48 45.1 & +22 08 06 & 0.1437 & 4 & 43080(300)
& 43206(302) & 637(5) & 2.070(166) & 2.36(26) & 12.10 \\
07163+0817 & 07 16 23.7 & +08 17 34 & 0.1107 & 1 & 33183(110)
& 33367(115) & 482(2) & 0.891(\phn89) & 1.37(11) & 11.53 \\
07572+0533 & 07 57 17.9 & +05 33 16 & 0.1894 & 1 & 56783(122)
& 57022(126) & 865(2) & 0.955(\phn76) & 1.30(20) & 12.04 \\
08201+2801 & 08 20 10.1 & +28 01 19 & 0.1680 & 5 & 50365(\phn70)
& 50583(\phn77) & 757(1) & 1.171(\phn70) & 1.43(16) & 12.00 \\
08279+0956 & 08 27 56.1 & +09 56 41 & 0.2085 & 1 & 62521(107)
& 62788(110) & 963(2) & 0.586(\phn64) & \<\phn1.26 & 11.75--12.01 \\
08449+2332 & 08 44 55.6 & +23 32 12 & 0.1510 & 1 & 45277(102)
& 45530(106) & 675(2) & 0.867(\phn69) & 1.20(17) & 11.79 \\
08474+1813 & 08 47 28.3 & +18 13 14 & 0.1450 & 5 & 43470(\phn70)
& 43739(\phn75) & 646(1) & 1.279(115) & 1.54(18) & 11.91 \\
09039+0503 & 09 03 56.4 & +05 03 28 & 0.1250 & 5 & 37474(\phn70)
& 37781(\phn73) & 551(1) & 1.484(\phn89) & 2.06(21) & 11.86 \\
09531+1430 & 09 53 08.3 & +14 30 22 & 0.2151 & 1 & 64494(148)
& 64818(149) & 998(2) & 0.777(\phn62) & 1.04(14) & 12.08 \\
09539+0857 & 09 53 54.9 & +08 57 23 & 0.1290 & 5 & 38673(\phn70)
& 39008(\phn72) & 570(1) & 1.438(101) & 1.04(18) & 11.79 \\
10035+2740 & 10 03 36.7 & +27 40 19 & 0.1662 & 1 & 49826(300)
& 50116(301) & 750(5) & 1.144(126) & 1.63(161) & 12.01 \\
10339+1548 & 10 33 58.1 & +15 48 11 & 0.1965 & 1 & 58906(122)
& 59242(123) & 903(2) & 0.977(\phn59) & 1.35(16) & 12.09 \\
10378+1108\tablenotemark{a} & 10 37 49.1 & +11 09 08 & 0.1362 & 2 & 40843(\phn61)
& 41190(\phn62) & 605(1) & 2.281(137) & 1.82(18) & 12.05 \\
11028+3130 & 11 02 54.0 & +31 30 40 & 0.1990 & 5 & 59659(\phn70)
& 59948(\phn73) & 915(1) & 1.021(\phn72) & 1.44(16) & 12.13 \\
11180+1623\tablenotemark{b}& 11 18 06.7 & +16 23 16 & 0.1660 & 5 & 49766(\phn70)
& 50104(\phn71) & 749(1) & 1.189(\phn95) & 1.60(18) & 12.02 \\

11524+1058 & 11 52 29.6 & +10 58 22 & 0.1784 & 1 & 53479(134)
& 53823(135) & 811(2) & 0.821(\phn66) & 1.17(15) & 11.93 \\
12005+0009 & 12 00 30.2 & +00 09 24 & 0.1226 & 1 & 36759(177)
& 37116(177) & 540(3) & 0.736(\phn88) & 0.98(15) & 11.52 \\
12018+1941\tablenotemark{a} & 12 01 51.8 & +19 41 46 & 0.1686 & 6 & 50559(\phn65)
& 50880(\phn67) & 762(1) & 1.761(123) & 1.78(23) & 12.16 \\
12032+1707 & 12 03 14.9 & +17 07 48 & 0.2170 & 5 & 65055(\phn70)
& 65382(\phn72) & 1008(1) & 1.358(\phn95) & 1.54(19) & 12.31 \\
12162+1047 & 12 16 13.9 & +10 47 58 & 0.1465 & 1 & 43931(149)
& 44267(150) & 654(2) & 0.725(\phn58) & \<\phn0.95 & 11.50--11.68 \\
12549+2403 & 12 54 53.4 & +24 03 57 & 0.1317 & 1 & 39491(145)
& 39772(147) & 582(2) & 0.739(\phn66) & 1.03(13) & 11.60 \\
13218+0552 & 13 21 48.4 & +05 52 40 & 0.2051 & 6 & 61488(\phn58)
& 61788(\phn62) & 946(1) & 1.174(\phn82) & 0.71(14) & 12.13 \\
14043+0624 & 14 04 20.0 & +06 24 48 & 0.1135 & 1 & 34025(114)
& 34283(117) & 496(2) & 0.795(\phn64) & 1.31(16) & 11.51 \\
14059+2000 & 14 05 56.4 & +20 00 42 & 0.1237 & 1 & 37084(\phn89)
& 37316(\phn93) & 543(1) & 0.857(120) & 1.88(32) & 11.68 \\
14070+0525\tablenotemark{a} & 14 07 00.3 & +05 25 40 & 0.2655 & 1 & 79591(400)
& 79847(401) & 1264(7) & 1.447(\phn87) & 1.82(18) & 12.55 \\
14553+1245 & 14 55 19.1 & +12 45 21 & 0.1249 & 1 & 37449(133)
& 37636(136) & 549(2) & 0.888(\phn53) & 1.17(16) & 11.62 \\
14586+1432 & 14 58 41.6 & +14 31 53 & 0.1477 & 1 & 44287(118)
& 44467(122) & 658(2) & 0.569(\phn91) & 1.07(17) & 11.64 \\
15224+1033 & 15 22 27.4 & +10 33 17 & 0.1348 & 1 & 40405(155)
& 40559(158) & 595(2) & 0.737(\phn74) & 0.72(15) & 11.57 \\
15587+1609 & 15 58 45.5 & +16 09 23 & 0.1375 & 1 & 41235(195)
& 41329(198) & 607(3) & 0.740(\phn52) & 0.82(21) & 11.60 \\
16100+2527 & 16 10 00.4 & +25 28 02 & 0.1310 & 3 & 39272(250)
& 39338(252) & 575(4) & 0.715(\phn50) & \<\phn1.38 & 11.39--11.63 \\
16255+2801 & 16 25 34.0 & +28 01 32 & 0.1340 & 1 & 40186(122)
& 40226(127) & 590(2) & 0.885(\phn88) & 1.26(26) & 11.69 \\
16300+1558 & 16 30 05.6 & +15 58 02 & 0.2417 & 6 & 72467(\phn64)
& 72515(\phn73) & 1133(1) & 1.483(134) & 1.99(32) & 12.47 \\
17161+2006 & 17 16 05.8 & +20 06 04 & 0.1098 & 1 & 32928(113)
& 32903(118) & 475(2) & 0.632(\phn44) & \<\phn1.37 & 11.16--11.43 \\
17539+2935 & 17 54 00.1 & +29 35 50 & 0.1085 & 6 & 32525(\phn58)
& 32441(\phn67) & 467(1) & 1.162(\phn58) & 1.36(19) & 11.58 \\
18368+3549 & 18 36 49.5 & +35 49 36 & 0.1162 & 2 & 34825(\phn40)
& 34688(\phn51) & 502(1) & 2.233(134) & 3.83(27) & 11.98 \\
18588+3517 & 18 58 52.4 & +35 17 04 & 0.1067 & 6 & 31973(\phn35)
& 31810(\phn46) & 458(1) & 1.474(103) & 1.75(33) & 11.66 \\
20248+1734 & 20 24 52.3 & +17 34 24 & 0.1208 & 6 & 36219(\phn87)
& 35943(\phn90) & 522(1) & 0.743(\phn82) & 2.53(38) & 11.68 \\
20286+1846 & 20 28 39.9 & +18 46 37 & 0.1347 & 1 & 40396(127)
& 40117(129) & 588(2) & 0.925(\phn74) & 2.25(16) & 11.81 \\
20450+2140 & 20 45 00.1 & +21 40 03 & 0.1284 & 1 & 38480(111)
& 38189(113) & 557(2) & 0.725(\phn51) & 1.90(15) & 11.67 \\
21077+3358 & 21 07 45.9 & +33 58 05 & 0.1764 & 1 & 52874(117)
& 52587(119) & 791(2) & 0.885(\phn88) & \<\phn1.55 & 11.75--11.98 \\
21272+2514 & 21 27 15.1 & +25 14 39 & 0.1508 & 1 & 45208(120)
& 44890(121) & 664(2) & 1.075(118) & \<\phn1.63 & 11.69--11.89 \\
22055+3024 & 22 05 33.6 & +30 24 52 & 0.1269 & 1 & 38041(\phn24)
& 37715(\phn29) & 550(0) & 1.874(356) & 2.32(23) & 11.93 \\
22116+0437 & 22 11 38.6 & +04 37 29 & 0.1939 & 1 & 58144(118)
& 57787(118) & 878(2) & 0.916(\phn73) & \<\phn1.03 & 11.85--12.01 \\
23019+3405 & 23 01 57.3 & +34 05 27 & 0.1080 & 6 & 32389(\phn28)
& 32061(\phn32) & 462(0) & 1.417(\phn99) & 2.11(38) & 11.69 \\
23028+0725 & 23 02 49.2 & +07 25 35 & 0.1496 & 1 & 44845(198)
& 44476(198) & 658(3) & 0.914(100) & \<\phn1.37 & 11.61--11.81 \\
23129+2548 & 23 12 54.4 & +25 48 13 & 0.1790 & 5 & 53663(\phn70)
& 53314(\phn71) & 803(1) & 1.811(145) & 1.64(44) & 12.21 \\
23199+0123 & 23 19 57.7 & +01 22 57 & 0.1367 & 3 & 40981(250)
& 40614(250) & 596(4) & 0.627(\phn63) & 1.03(16) & 11.58 \\
23234+0946 & 23 23 23.6 & +09 46 15 & 0.1279 & 6 & 38356(\phn24)
& 37988(\phn24) & 554(0) & 1.561(\phn94) & 2.11(30) & 11.88 \\
\enddata
\tablenotetext{a}{A known OHM included in the survey sample.}
\tablenotetext{b}{{\it IRAS} 11180+1623 is not in the PSCz catalog (excluded 
by the PSCz mask; Saunders \etal 2000).  It is in the 1 Jy survey (Kim \& Sanders 1998).}
\tablerefs{Redshifts were obtained from:  
(1) Saunders \etal 2000; 
(2) Strauss \etal 1992; 
(3) Lawrence \etal 1999; 
(4) Lu \& Freudling 1995;
(5) Kim \& Sanders 1998; 
(6) Fisher \etal 1995.}
\end{deluxetable}

\begin{deluxetable}{lcrcccrccr}
\tabletypesize{\scriptsize}
\tablecaption{Arecibo Survey OH Megamasers:  OH Line and 1.4 GHz Continuum Properties 
	\label{detectOH}}
\tablewidth{0pt}
\tablehead{
\colhead{{\it IRAS} Name} &  
\colhead{$cz_{1667,\odot}$} & 
\colhead{$f_{1667}$} & 
\colhead{$W_{1667}$} &
\colhead{$\Delta \nu_{1667}$\tablenotemark{a}} & 
\colhead{$\Delta \mbox{v}_{1667}$\tablenotemark{b}} & 
\colhead{$R_H$} & 
\colhead{$\log L_{FIR}$} & 
\colhead{$\log L_{OH}$} &
\colhead{$f_{1.4GHz}$\tablenotemark{c}}  \\
\colhead{FSC} &\colhead{km/s} & \colhead{mJy} & \colhead{MHz}
& \colhead{MHz} & \colhead{km/s} &\colhead{} & \colhead{$h^{-2}_{75} L_\odot$} 
& \colhead{$h^{-2}_{75} L_\odot$} & \colhead{mJy} \\
\colhead{(1)}&\colhead{(2)}&\colhead{(3)}&\colhead{(4)}&\colhead{(5)}&
\colhead{(6)}&\colhead{(7)}&\colhead{(8)}&\colhead{(9)}&\colhead{(10)}
}
\startdata
01562+2528 & 49814(15) & 6.95 & 1.29 & 1.04 & 218 & 5.9 & 11.90 & 3.25 & 6.3(0.5) \\
02524+2046 & 54162(15) & 39.82 & 0.50 & 0.36 & \phn76 & 3.2 & 11.81--12.28 & 3.74 & 2.9(0.6)\\
03521+0028 & 45512(14) & 2.77 & 0.61 & 0.29 & \phn59 & 5.8 & 12.28 & 2.44 & 6.7(0.6) \\
03566+1647 & 39865(14) & 1.96 & 0.98 & 0.23 & \phn48 & $\gtrsim\phn9.6$ & 11.40--11.75 & 2.32 & 3.5(0.5) \\
04121+0223 & 36590(14) &  2.52 & 0.76 & 1.04 & 209 & 2.9 & 11.40--11.69 & 2.32 & 3.1(0.5)\\
06487+2208 & 42972(14) & \phn6.98 & 0.66 & 0.37 & \phn76 & 8.0 & 12.10 & 2.82 & 10.8(0.6) \\
07163+0817 & 33150(14) &  4.00 & 0.69 & 0.12 & \phn24 & $\sim\phn5.5$ & 11.53 & 2.37 & 3.5(0.5)\\
07572+0533 & 56845(15) &  2.26 & 1.03 & 0.73 & 156 & $\geq10.4$ & 12.04 & 2.74 & $< 5.0$\phn\\
08201+2801 & 50325(15) & 14.67 & 0.97--1.19 & 0.98 & 205 & $\geq\phn8.2$ & 12.00 & 3.45 & 16.7(0.7)\\
08279+0956 & 62422(15) &  4.79 & 1.02 & 0.95 & 207 & 5.9 & 11.75--12.01 & 3.23 & 4.4(0.8)\\
08449+2332 & 45424(14) &  2.49 & 1.09 & 0.47 & \phn97 & $\geq11.0$ & 11.79 & 2.59 & 6.1(0.5)\\
08474+1813 & 43750(14) &  2.20 & 1.29--1.70 & 1.98 & 409 & $\geq\phn3.0$ & 11.91 & 2.70 & 4.2(0.5)\\
09039+0503 & 37720(14) &  5.17 & 1.23 & 1.05 & 212 & 8.5 & 11.86 & 2.83 & 6.6(0.5)\\
09531+1430 & 64434(15) &  3.98 & 1.03 & 1.17 & 256 & $\sim\phn3.4$ & 12.08 & 3.42 & 3.0(0.5)\\
09539+0857 & 38455(14) & 14.32 & 1.47 & 1.56 & 317 & 2.5 & 11.79 & 3.48 & 9.5(1.2)\\
10035+2740 & 50065(14) &  2.29 & 0.74 & 0.31 & \phn65 & $\sim15.4$ & 12.01 & 2.50 & 6.3(0.5)\\
10339+1548 & 58983(15) &  6.26 & 0.28 & 0.19 & \phn40 & $\gtrsim14.5$ & 12.09 & 2.65 & 5.1(0.5)\\
10378+1108\tablenotemark{d} & 40811(14) & 19.70 & 1.50 & 0.87 & 177 & \nodata & 12.05 & 3.54 & 8.9(0.6)\\
11028+3130 & 59619(15) &  4.27 & 0.72 & 0.41 & \phn89 & 5.5 & 12.13 & 2.97 & $< 5.0$\phn\\
11180+1623\tablenotemark{e}& 49783(14) &  1.82 & 0.42 & 0.61 & 127 & $\geq\phn5.1$ & 12.02 & 2.34 & 4.2(0.5) \\
11524+1058 & 53404(15) &  3.17 & 1.21 & 1.32 & 279 & $\sim\phn4.9$ & 11.93 & 2.98 & $< 5.0$\phn\\
12005+0009 & 36472(14) &  3.51 & 0.71 & 0.41 & \phn82 & $\sim\phn2.0$ & 11.52 & 2.62 & 5.4(0.6) \\
12018+1941\tablenotemark{d} & 50317(15) & 2.63 & 0.81 & 0.86 & 181 & $\geq\phn5.6$ & 12.16 & 2.59 & 6.5(0.5)\\
12032+1707 & 64920(15) & 16.27 & 2.69 & 3.90 & 853 & \nodata & 12.31 & 4.15 & 28.7(1.0)\\
12162+1047 & 43757(14) &  2.07 & 0.64 & 0.51 & 105 & $\geq11.1$ & 11.50--11.68 & 2.25 & 6.8(1.6) \\
12549+2403 & 39603(14) &  1.79 & 0.89 & 0.50 & 102 & $\sim\phn2.6$ & 11.60 & 2.37 & 3.7(0.5) \\
13218+0552 & 61268(15) &  4.01 & 2.49 & 1.45 & 314 & \nodata & 12.13 & 3.45 & 5.3(0.5) \\
14043+0624 & 33912(14) &  2.75 & 0.33 & 0.27 & \phn54 & 1.4 & 11.51 & 2.10 & 15.6(1.0)\\
14059+2000 & 37246(14) & 15.20 & 1.10 & 0.80 & 161 & 5.3 & 11.68 & 3.34 & 7.5(0.5) \\
14070+0525\tablenotemark{d} & 79929(16) & 8.37 & 3.21 & 2.62 & 596 & \nodata & 12.55 & 4.13 & 4.0(0.6)\\
14553+1245 & 37462(14) &  2.93 & 0.39 & 0.38 & \phn77 & $\geq14.5$ & 11.62 & 2.27 & 3.8(0.5) \\
14586+1432 & 44380(14) & 7.11 & $\leq2.67$ & 1.79 & 369 & \nodata & 11.64 & 3.41 & 11.1(0.6)\\
15224+1033 & 40290(14) & 12.27 & 0.73--0.80 & 0.15 & \phn31 & $\geq\phn9.5$ & 11.57 & 3.04 & 3.6(0.5)\\
15587+1609 & 40938(14) & 13.91 & 0.99 & 0.86 & 176 & 6.9 & 11.60 & 3.26 & $< 5.0$\phn\\
16100+2527 & 40040(14) &  2.37 & 0.60 & 0.23 & \phn46 & 3.2 & 11.39--11.63 & 2.29 & $< 5.0$\phn\\
16255+2801 & 40076(14) &  7.02 & 0.45 & 0.39 & \phn79 & $\geq13.7$ & 11.69 & 2.57 & $< 5.0$\phn\\
16300+1558 & 72528(15) & \phn3.12 & 0.56 & 0.59 &    131 & \nodata & 12.47 & 2.85 & 7.9(0.5) \\
17161+2006 & 32762(14) & 4.84 & 0.62 & 0.38 & \phn76 & $\sim\phn6.2$ & 11.16--11.43 & 2.39 & 7.3(0.6)  \\
17539+2935 & 32522(14) & \phn0.76 & 0.72 & 0.81 &    161 & $\geq\phn2.9$ & 11.58 & 1.76 & 4.0(0.6) \\
18368+3549 & 34832(14) & \phn4.58 & 1.79 & 2.10 &    421 & $\sim\phn$9.5 & 11.98 & 2.85 & 21.0(0.8)\\
18588+3517 & 31686(14) & \phn7.37 & 0.56 & 0.32 & \phn64 & 5.1 & 11.66 & 2.52 & 5.9(0.5) \\
20248+1734 & 36538(14) & \phn2.61 & 1.36 & 0.88 &    177 & $\sim\phn$6.8 & 11.68 & 2.53 & $<5.0$\phn \\
20286+1846 & 40471(14) &    15.58 & 1.51 & 1.10 &    224 & $\geq\phn4.4$ & 11.81 & 3.41 & $<5.0$\phn\\
20450+2140 & 38398(14) & \phn2.27 & 0.67 & 0.71 &    144 & $\geq\phn6.2$ & 11.67 & 2.24 & 5.0(0.5) \\
21077+3358 & 52987(15) & \phn5.04 & 1.86 & 1.15 &    243 & $\geq\phn7.4$ & 11.75--11.98 & 3.26 & 9.4(1.0) \\
21272+2514 & 45032(14) &    16.33 & 1.87 & 1.27 &    263 & 13.7 & 11.69--11.89 & 3.66 & 4.4(0.5) \\
22055+3024 & 37965(14) &  6.35 & 0.77 & 0.46 & \phn92 & 6.2 & 11.93 & 2.73 & 6.4(0.5)\\
22116+0437 & 58180(15) & \phn1.76 & 1.16 & 0.56 & 121 & $\sim\phn$5.2 & 11.85--12.01 & 2.77 & 8.4(0.6) \\
23019+3405 & 32294(14) &  3.58 & 0.52 & 0.28 & \phn57 & $\geq15.6$ & 11.69 & 2.12 & 7.7(0.5)\\
23028+0725 & 44529(14) &  8.69 & 1.09 & 1.06 & 219 & 1.9 & 11.61--11.81 & 3.29 & 19.5(1.1)\\
23129+2548 & 53394(15) &  4.59 & 2.0\phn & 1.78 & 376 & \nodata & 12.21 & 3.27 & 4.7(0.5)\\
23199+0123 & 40680(14) &  1.80 & 0.82 & 0.68 & 139 & $\sim\phn2.3$ & 11.58 & 2.38 & 3.0(0.5)\\
23234+0946 & 38240(14) &  3.32 & 1.23 & 1.32 & 266 & 2.4 & 11.88 & 2.75 & 11.6(1.0)\\
\enddata
\tablenotetext{a}{The frequency width $\Delta \nu_{1667}$ is the {\it observed} FWHM.}
\tablenotetext{b}{The velocity width $\Delta \mbox{v}_{1667}$ is the {\it rest frame} FWHM.  The rest
frame and observed widths are related by 
$\Delta \mbox{v}_{rest} = c(1+z)(\Delta\nu_{obs}/\nu_\circ)$.}
\tablenotetext{c}{1.4 GHz continuum fluxes are courtesy of the NRAO VLA Sky Survey 
(Condon \etal 1998).}
\tablenotetext{d}{Redetection of a known OHM included in the survey sample.}
\tablenotetext{e}{{\it IRAS} 11180+1623 is not in the PSCz catalog (excluded 
by the PSCz mask; Saunders \etal 2000).  
It is in the 1 Jy survey (Kim \& Sanders 1998).}
\end{deluxetable}

\subsection{OH Megamaser Detections}\label{subsec:detections}

Tables \ref{detectFIR} and \ref{detectOH} list respectively the 
optical/FIR and radio properties of the 50 new OHM detections and 
3 redetections.  Note that OHM {\it IRAS} F11180+1623
is {\it not} in the PSCz sample, but was observed
along with other OHM candidates not found in the PSCz sample to fill in 
telescope time when local sidereal time coverage of the official sample 
was sparse.  This
detection is not included in any of the survey statistics or interpretation,
including the OH LF\@.  Spectra of the 53 OHMs appear in Papers I, II, and 
III\@. Table \ref{detectFIR} lists the optical redshifts
and FIR properties of the non-detections in the following format:

\noindent
Column (1):  {\it IRAS} Faint Source Catalog (FSC) name.  

\noindent
Columns (2) and (3):  Source coordinates (epoch B1950.0) 
from the FSC, or the Point Source Catalog (PSC) if unavailable in the FSC.  

\noindent
Columns (4), (5) and (6):  Heliocentric optical redshift, 
reference, and corresponding
velocity.  Uncertainties in velocities are listed whenever they are 
available.  

\noindent
Column (7):  Cosmic microwave background rest-frame velocity.  This is
computed from the heliocentric velocity using the solar motion with respect
to the CMB measured by Lineweaver \etal (1996):  
$c z_\odot = 368.7 \pm 2.5$ km s\minusone\
towards $(l,b) = (264\fdg31 \pm 0\fdg16 , 48\fdg05 \pm 0\fdg09)$.

\noindent
Column (8):  Luminosity distance computed from $z_{CMB}$ via 
\begin{equation}
D_L = (1+z_{CMB}){c \over H_\circ}\int_0^{z_{CMB}} 
 	\left[(1+z^\prime)^3\Omega_M
		+ \Omega_\Lambda\right]^{-{1\over2}}\,dz^\prime, 
							\label{eqn:D_L_z}
\end{equation}
assuming $\Omega_M = 0.3$ and $\Omega_\Lambda = 0.7$.  

\noindent
Columns  (9) and (10):  {\it IRAS} 60 and 100 $\mu$m flux 
densities in Jy.  FSC flux densities are listed whenever they are 
available.  Otherwise, PSC flux densities
are used.  Uncertainties refer to the last digits of each measure, and upper 
limits on 100 $\mu$m flux densities are indicated by a ``less-than'' symbol. 

\noindent
Column (11):  The logarithm of the far-infrared luminosity in units
of $h_{75}^{-2} L_\odot$.  
\LFIR\ is computed following the prescription of Fullmer \& Lonsdale (1989):  
\LFIR$ = 3.96\times 10^5 D_L^2 (2.58 f_{60} + f_{100})$, 
where $f_{60}$ and $f_{100}$ are the 60 and 100 
$\mu$m flux densities expressed in Jy, $D_L$ is in $h_{75}^{-1}$ Mpc, 
and \LFIR\ is in units of $h_{75}^{-2} L_\odot$.  
If $f_{100}$ is only available as an upper limit, the permitted range
of \LFIR\ is listed.  The lower bound on \LFIR\ is computed for $f_{100}=0$ Jy,
and the upper bound is computed with $f_{100}$ set equal to its upper limit.
The uncertainties in $D_L$ and in the {\it IRAS} flux densities 
typically produce an uncertainty in $\log L_{FIR}$ of $0.03$.  

Table \ref{detectOH} lists the OH emission
properties and 1.4 GHz flux density of the OH detections in the following
format:

\noindent
Column (1):  {\it IRAS} FSC name.

\noindent
Column (2):  Measured heliocentric velocity of the 1667.359 MHz 
line, defined by the center of the FWHM of the line.  The uncertainty in the
velocity of the line center is estimated assuming an uncertainty of $\pm 1$
channel ($\pm 49$ kHz) on each side of the line.  
Although one can generally
determine emission line centers with much higher precision when the shapes
of lines are known, OHM line profiles are asymmetric, multi-component, 
and non-gaussian (see Papers I--III).  They defy simple 
shape descriptions, so we use this conservative and basic prescription 
to quantify the uncertainty in the line centers.

\noindent
Column (3):  Peak flux density of the 1667 MHz OH line in mJy.

\noindent
Column (4):  Equivalent width-like measure in MHz.  
$W_{1667}$ is the ratio of the integrated 1667 MHz line flux to its 
peak flux.  Ranges are listed for $W_{1667}$ in cases where the identification
of the 1665 MHz line is unclear, but in many cases the entire emission 
structure is included in $W_{1667}$ as indicated in the discussion of 
each source in Papers I--III.

\noindent
Column (5):  Observed FWHM of the 1667 MHz OH line in MHz.

\noindent
Column (6):  Rest frame FWHM of the 1667 MHz OH line in km 
s\minusone.
The rest frame width was calculated from the observed width as
$\Delta \mbox{v}_{rest} = c (1+z) (\Delta \nu_{obs} / \nu_\circ)$.

\noindent
Column (7):  Hyperfine ratio, defined by $R_H = F_{1667}/F_{1665}$, where 
$F_\nu$ is the integrated flux density across the emission line centered on 
$\nu$.  $R_H = 1.8$ in thermodynamic equilibrium.  In many cases, the 1665 MHz
OH line is not apparent, or is blended into the 1667 MHz OH line, and a good
measure of $R_H$ becomes difficult without a model for the line profile.  It is
also not clear that the two lines should have similar profiles, particularly if the
lines are aggregates of many emission regions in different saturation states. 
Some spectra allow a lower limit to be placed on $R_H$, indicated by a 
``greater than''
symbol.  Blended or noisy lines have uncertain values of $R_H$, and are indicated 
by a tilde, but in some cases, separation of the two OH lines is impossible
and no value is listed for $R_H$.  

\noindent
Column (8):  Logarithm of the FIR luminosity, as in Table \ref{detectFIR}.

\noindent
Column (9):  Logarithm of the measured isotropic OH line luminosity, 
which includes the 
integrated flux density of both the 1667.359 and the 1665.4018 MHz 
lines.

\noindent
Column (10):  1.4 GHz continuum fluxes from the NRAO VLA Sky Survey
(Condon \etal 1998).
If no continuum source lies within 30\arcsec\ of the {\it IRAS} 
coordinates, an upper limit of 5.0 mJy is listed.

\section{The OH Luminosity Function}\label{sec:OHLF}

\subsection{Computing a Spectral Line Luminosity Function}\label{subsec:lineLF}

The luminosity function $\Phi(L)$ is the number density of objects
with luminosity $L$ per (logarithmic) interval in $L$.  An unbiased 
direct measurement
of $\Phi(L)$ would require that all objects with a given luminosity be
detected within the survey volume, which is generally not possible in 
a flux-limited survey.  Instead, each object in a survey has an 
effective volume in which it could have been detected by the survey, and 
the sum of detections weighted by their available volumes $V_a$ determines 
the luminosity function.  
The most general unbiased maximum likelihood computation of a 
luminosity function is computed from $V_a$ following the prescription 
\citep{page00}:
\begin{equation}
	\Phi(L) = {1 \over \Delta\log L} \sum {1 \over V_a(L)}
\end{equation}
where the sum is over the detected objects in the luminosity bin 
$\Delta\log L$ centered on $L$ and 
redshift bin $\Delta z$ centered on $z$.
The uncertainty in the luminosity function is 
\begin{equation}
	\sigma_\Phi = {1\over\Delta\log L}\left(\sum {1\over V_a^2}\right)^{1\over2}.
\end{equation}
Computation of the volume available to each detection depends on the
areal coverage of the survey $\Omega$ and the maximum detectable distance
of each object detected ($D_{L,max}$ or $z_{max}$).   
The relationship between physical comoving volume and luminosity distance is
given by 
\begin{equation}
	V = {\Omega\over3}\left(D_L \over 1+z\right)^3 \label{eqn:V}
\end{equation}
{\it only} when $\Omega_k = 0$, where 
$\Omega_k = 1- \Omega_M - \Omega_\Lambda$ \citep{wei72}.
From the definition of luminosity distance,
\begin{equation}
	D_L \equiv \left(L \over 4\pi S\right)^{1\over2}
						\label{eqn:D_L} 
\end{equation}
for $\Omega_k = 0$, where $L$ is the integrated line luminosity and
S is the integrated line flux.  
Detection of spectral lines depends on the peak line flux density 
rather than the integrated line flux.  The luminosity distance of 
Equation \ref{eqn:D_L} must be modified by a factor $1+z$ in order
to change to rest-frame luminosity density and flux density:
\begin{equation}
 	D_L = \left( L_{\nu_\circ} \over 4\pi S_\nu\right)^{1\over2} 
		(1+z)^{1\over2}.
\end{equation}
For a survey with sensitivity limit $n\sigma_\nu$, where $\sigma_\nu$
is the RMS noise flux density at frequency $\nu$, and assuming that 
this noise is independent of frequency, $L_{\nu_\circ}$ 
is an invariant quantity which relates the observed luminosity distance
and peak flux density $S_\nu$ to the maximum detectable distance and 
the sensitivity limit $n\sigma_\nu$.  
We thus obtain an expression for the maximum detectable distance, 
$D_{L,max}$, which depends on itself through $z_{max}$ 
\begin{equation}
	D_{L,max} = D_L \left( S_\nu \over n\sigma_\nu \right)^{1\over2}
			\left( 1+z_{max} \over 1+z \right)^{1\over2}
						\label{eqn:D_Lmax1}
\end{equation}
and must be solved such that $D_{L,max}$ can be related to observed
quantities only.  Equation \ref{eqn:D_L_z} has no simple analytic  
solution for $\Omega_\Lambda \neq 0$ and must be solved numerically.
Inserting this solution into Equation \ref{eqn:D_Lmax1}, we
obtain a relation in $z_{max}$:
\begin{eqnarray}
 \lefteqn{\int_0^{z_{max}} \left[(1+z^\prime)^3\Omega_M
		+ \Omega_\Lambda\right]^{-{1\over2}}\,dz^\prime} \nonumber \\
 & & {\mbox{} = {H_\circ D_L \over c \sqrt{1+z}}\left(S_\nu\over n\sigma_\nu\right)^{1\over2}
	(1+z_{max})^{-{1\over2}}}
\end{eqnarray}
which has no analytical solution.  
%
%
The numerical solution for $z_{max}$ (and the equivalent $D_{L,max}$) 
can be inserted
into an expression for $V_{max}$ obtained from Equation \ref{eqn:V}.
Finally, we obtain the volume available to a given emission line source in 
the survey in the case where there is a minimum survey redshift $z_{min}$:
\begin{eqnarray}
V_a &=& V_{max} - V_{min} \nonumber\\
    &=& {\Omega\over3}\left[\left(D_{L,max}\over1+z_{max}\right)^3
	- \left(D_{L,min}\over1+z_{min}\right)^3\right].
\end{eqnarray}
Note that if $z_{max}$ is greater than the upper bound in redshift of 
the survey, then we set $z_{max}$ equal to this upper bound.

\subsection{Computing a Continuum Luminosity Function}\label{subsec:ctnmLF}

The procedure for computing a continuum luminosity function closely follows
the steps for the emission line computation.  The detection threshold is
determined in the same manner, from a flux density sensitivity rather
than an integrated flux.  
Unlike spectral line surveys, however, continuum surveys do not 
usually tune the bandpass to the redshift of each observed object.  This
introduces an error in computing the invariant luminosity from the
measured flux because a different portion of the rest-frame spectral
energy distribution is sampled for each object due to the redshift.  A
k-correction must be applied to the measured flux density to obtain the
correct luminosity and hence to compute the maximum detectable distance
for each source.  The net k-correction between $z$ and $z_{max}$ will 
modify Equation \ref{eqn:D_Lmax1} slightly
\begin{equation}
	D_{L,max} = D_L \left( S_\nu \kappa \over n\sigma_\nu \right)^{1\over2}
			\left( 1+z_{max} \over 1+z \right)^{1\over2},
\end{equation}
but will add another wrinkle to the derivation of $V_{max}$ 
in \S \ref{subsec:lineLF} because the k-correction will itself 
depend on $D_L$.  We will approximate
the k-correction to be constant at first, determine $D_{L,max}$,
recompute the k-correction, and then correct $D_{L,max}$ if necessary.
The k-correction is a weak function of $D_L - D_{L.max}$, 
and this single iteration approach to computing $D_{L,max}$ is adequate.

\subsection{The OH Megamaser Luminosity Function}\label{subsec:OHLF}

\begin{figure}[b!]
\plotone{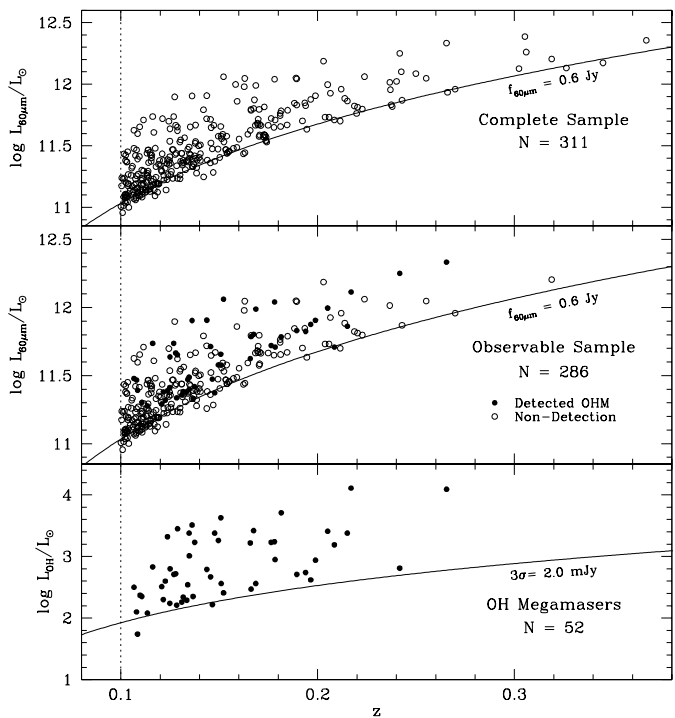}
\caption[The Arecibo OH Megamaser Survey Sample, Observable Candidates, 
and Detected OH Megamasers]
{The Arecibo OH megamaser survey sample, observable candidates, 
and detected OH Megamasers.  Shown are the 60 \micron\ luminosities 
and the redshifts of the complete sample of OHM candidates extracted from the
PSCz ({\it top}), the observable subset of these candidates which have unambiguous
OH line properties ({\it middle}), and the OH line luminosities of the detected
OHMs versus redshift ({\it bottom}).  The dotted vertical lines indicate the 
$z=0.1$ cutoff, and the solid lines indicate the flux density limits 
of the PSCz and the OHM survey.  
\label{survey_z}}
\end{figure}

The Arecibo OH megamaser survey is a flux-limited survey for OH emission
lines, but has three additional constraints:  (1) objects observed in the
survey are selected from a flux-limited 60 micron catalog (the PSCz; 
Saunders \etal 2000), (2) the survey has low and high redshift cutoffs
at $z=0.1$ and $z=0.23$, and (3) the survey cannot include objects close
to $z=0.174$ where the OH lines are redshifted into the strong Galactic \HI
emission.  The redshift cutoffs are imposed by radio 
frequency interference (RFI) encountered at Arecibo above 1510 MHz and 
below 1355 MHz.  The typical RMS flux density for a 12 minute integration
is 0.65 mJy, and detections are generally made at the $3\sigma$ level
or greater, which is 2 mJy.  These constraints are well-illustrated by
Figure \ref{survey_z}, which shows redshift cutoffs, the observable 
candidates, and the sensitivity limits in both OH line luminosity and 
FIR flux.  

Ignoring constraint (1) for now, 
computation of the OH LF can follow the prescription outlined in \S
\ref{subsec:lineLF} by setting the redshift bin $\Delta z$ to span
$z=0.1$ to $z=0.23$ with $z_{min}=0.1$.  We use $n\sigma_\nu = 2.0$ mJy
and can effectively ignore the thin shell of space centered on $z=0.174$
since its contribution to the total volume is negligible.  The survey
solid angle spans $0^\circ < \delta < 37^\circ$ such that 
$\Omega = 3.78$ steradians (30$\%$ of the sky).  

Now fold in the PSCz selection criteria.  First, the PSCz does not
completely cover the $0^\circ < \delta < 37^\circ$ band.  It excludes the
galactic plane and areas with inadequate or confused
{\it IRAS} coverage (Saunders \etal 2000).  
The PSCz mask excludes 18$\%$ of the Arecibo coverage,
reducing the survey solid angle to $\Omega = 3.21$ steradians (25.5$\%$ of
the sky).  The survey volume from $z=0.1$ to $z=0.23$ is thus 0.63 Gpc$^3$.  

Second, the PSCz has a 60 \micron\ flux density limit of 0.6 Jy.  Hence, the
volume available to each object in the survey can potentially be limited
by this cutoff rather than the OH line detectability.  The true volume
available to a given object in the survey is now 
\begin{equation}
	V_a = \min\{V_a^{OH},V_a^{60\mu m}\}.
\end{equation}
Computation of $V_a^{60\mu m}$ follows the continuum prescription outlined
in \S \ref{subsec:ctnmLF}, and the calculation of the LF uses this 
more restrictive definition of $V_a$.  The luminosity bins $\Delta\log L$
refer to the integrated OH line luminosities of the OHMs.  Hence, the
details of detecting OHMs from both surveys are folded into $V_a$, and
the luminosity intervals in the LF incorporate the information 
about the OH line luminosities.

Third, the {\it IRAS} 60 \micron\ flux measurements require net k-corrections
from $D_L$ to $D_{L,max}$.  
The k-corrections themselves are derived from spectral energy distribution
models for star forming galaxies developed by Dale \etal (2001).  The
models depend on the rest-frame FIR color $\log (f_{60\mu m}/f_{100\mu m})$
which is corrected from the observed color to the rest frame color
using values tabulated by D. Dale (2001, private communication).  The typical 
color correction from $z=0.15$ to $z=0$ for 
the OHM host sample is 
0.05--0.10 (colors get ``warmer'').  The net k-correction for an OHM 
detected at $z=0.15$ which is detectable to $z=0.20$ with rest frame
FIR color of $-0.10$ is $\kappa=0.98$.  
Although the OHM sample spans a wide range of FIR colors, the 
net k-corrections vary little from source to source in the range 
$z=0.1$--0.23.  Objects with no 100 \micron\
detection generally have a range of possible net k-corrections centered on 
unity, which is adopted for lack of better information.  

\begin{figure*}[!ht]
\epsscale{1.6}
\plotone{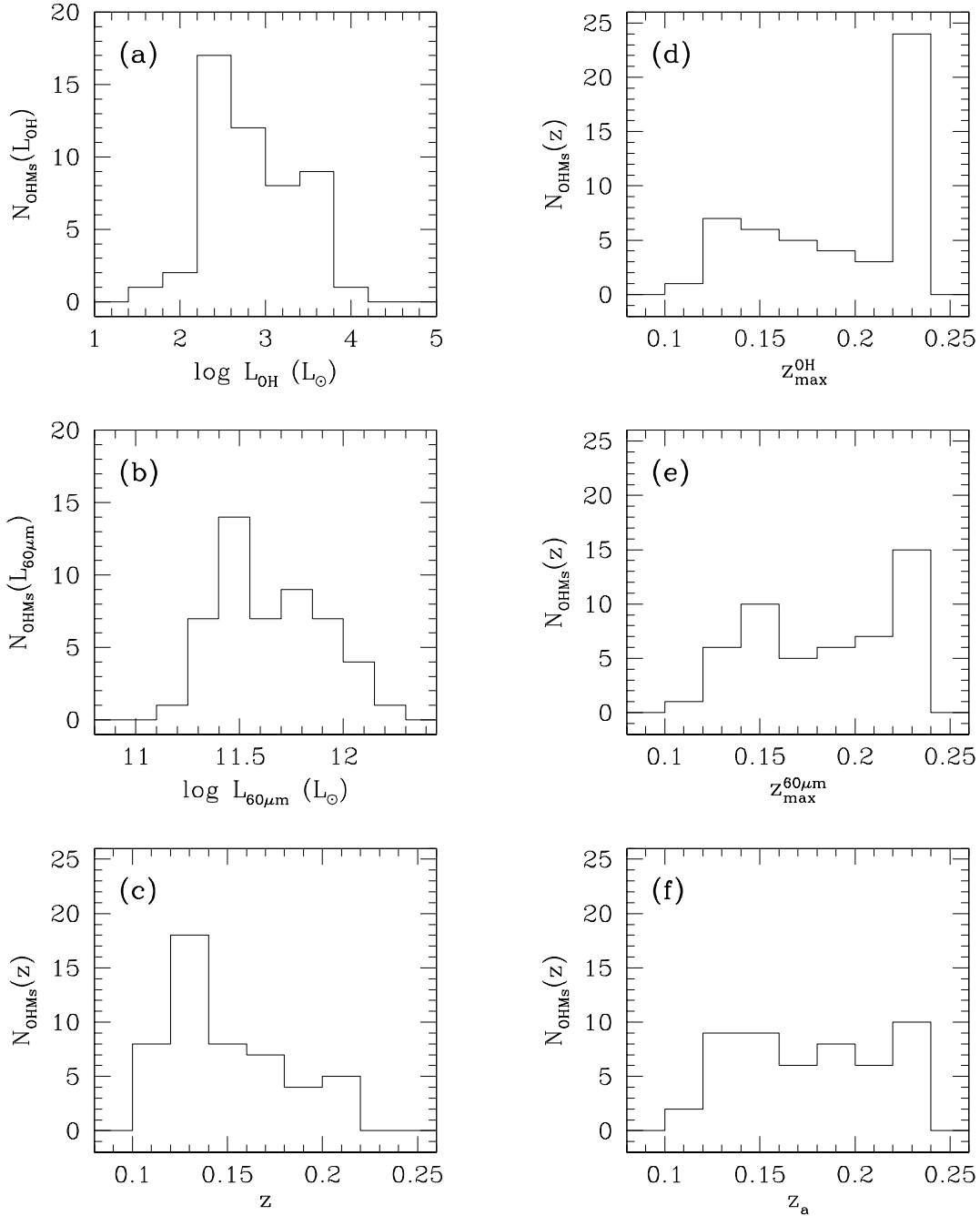}
\caption[The OH Megamaser Sample:  \LOH, $L_{60\mu m}$, Redshift, 
and Maximum Detectable Redshift Distributions]
{The OH megamaser sample:  \LOH, $L_{60\mu m}$, redshift, 
and maximum detectable redshift distributions.
Panels show: 
({\it a}) The \LOH\ distribution of the survey OHM detections; 
({\it b}) The $L_{60\mu m}$ distribution; 
({\it c}) The redshift distribution;
({\it d}) The maximum detectable redshift distribution calculated from
	the OH emission line;
({\it e}) The maximum detectable redshift distribution calculated from
	the 60 \micron\ flux density; and
({\it f}) The available redshift distribution computed from 
	$\min\{z_{max}^{OH},z_{max}^{60\mu m}\}$.
These distributions exclude two OHMs at $z>0.23$.
\label{hists1}}
\end{figure*}

Figure \ref{hists1} shows the OH luminosity, FIR luminosity, and 
redshift distributions of the OHM sample.  Also shown are the maximum
detectable redshift distributions computed from the OH detections, 
the 60 \micron\ detections, and the available redshift distribution
$z_a$.  Note that the $z_a$ distribution is quite flat but that the
maximum detectable redshift distribution for OH indicates two populations
of OHMs:  a population which is just detected by the survey and indicates a 
large population of OHMs which would be detected by a deeper survey, and
a population of ``over-luminous'' OHMs which can be detected in short 
integration times out to $z\simeq1$, similar to QSOs or radio galaxies.  
Nearly half of the OHM sample falls into this interesting latter category.
These are the objects which are useful for studies of galaxy evolution 
out to large redshifts.  
There is a hint of this population dichotomy in the plots of 
\LOH\ versus \LFIR\ analyzed in Paper III.

\begin{figure*}[ht!]
\epsscale{1.6}
\plotone{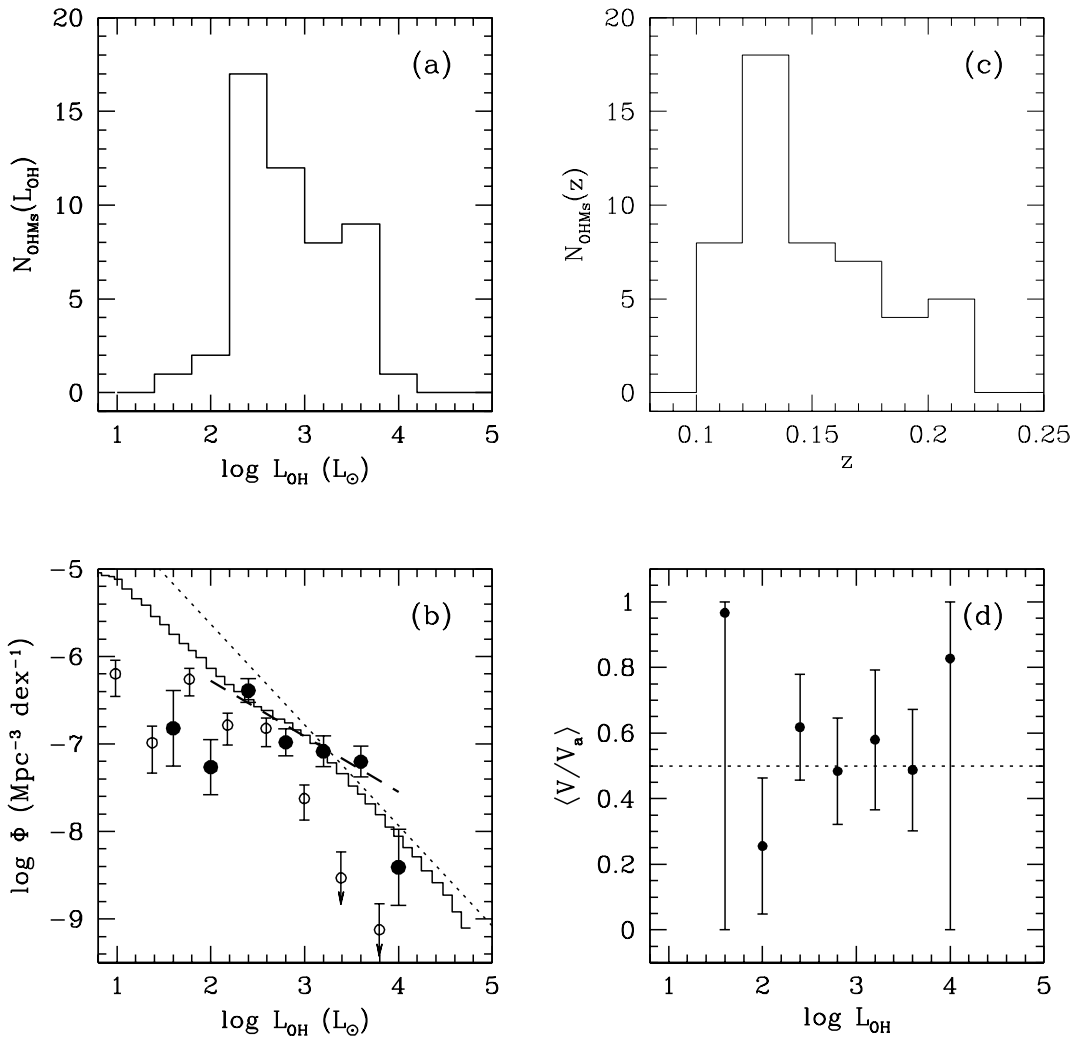}
\caption[The OH Megamaser Luminosity Function]
{The OH megamaser luminosity function. 
Panels show: 
({\it a}) The \LOH\ distribution of the survey OHM detections; 
({\it b}) The OH luminosity function ({\it filled circles}) and a power-law 
	fit to the well-sampled data points ($\Phi \propto L_{OH}^{-0.64}$; 
	{\it dashed line}), the OH LF derived by Baan (1991; {\it open circles}), and
	the OH LF computed by Briggs (1998) for a fixed and variable 
	OHM fraction in LIRGs ({\it dotted line and histogram, respectively});
({\it c}) The redshift distribution; and
({\it d}) The average ratio of $V$ to $V_a$ which is consistent with 
	0.5 in well-sampled bins (a uniformly distributed sample 
	has $\langle V/V_a\rangle = 0.5$).
These distributions and calculations exclude two OHMs at $z>0.23$.
\label{hists3}}
\end{figure*}

We combine all of the survey constraints and compute an OH LF 
which is presented in Figure \ref{hists3}.
A power-law fit to the well-sampled OH LF points gives 
\begin{equation}
	\Phi = (9.8^{+31.9}_{-7.5}\times10^{-6})\ L_{OH}^{-0.64\pm0.21}\ \
		\mbox{Mpc$^{-3}$ dex$^{-1}$}  	\label{eqn:LF}
\end{equation}
where \LOH\ is expressed in solar luminosities.
The uniformity of the sampling in space is checked
with the $\langle V/V_a \rangle$ test (Schmidt 1968).  A uniformly distributed
sample between 0 and 1 has mean 0.5.  Hence, the $\langle V/V_a \rangle$
values consistent with 0.5 in well-sampled \LOH\ bins shown in Figure 
\ref{hists3} indicate a uniformly distributed sample of OH megamasers.
Both the OH LF and $\langle V/V_a \rangle$ are tabulated in Table 
\ref{LF+ratio}, which includes the number of OHMs available in each \LOH\
bin.

\begin{deluxetable}{cccc}
\tabletypesize{\footnotesize}
\tablecaption{The OH Luminosity Function
	\label{LF+ratio}}
\tablewidth{0pt}
\tablehead{
\colhead{$\log L_{OH}$} & \colhead{N(OHMs)} & \colhead{$\Phi$}  
& \colhead{$\langle V/V_a \rangle$} \\
\colhead{$h^{-2}_{75} L_\odot$} 
& \colhead{} & \colhead{Mpc$^{-3}$ dex\minusone}   
}
\startdata
1.6 & \phn1 & $(1.5\pm1.5)\times10^{-7}$ & \phd\phd$0.97\pm$\nodata\phn \\
2.0 & \phn2 & $(5.4\pm3.9)\times10^{-8}$ & $0.26\pm0.21$ \\
2.4 & 17 & $(4.1\pm1.3)\times10^{-7}$ & $0.62\pm0.16$ \\
2.8 & 12 & $(1.0\pm0.4)\times10^{-7}$ & $0.48\pm0.16$ \\
3.2 & \phn8 & $(8.2\pm3.3)\times10^{-8}$ & $0.58\pm0.21$ \\
3.6 & \phn9 & $(6.3\pm2.5)\times10^{-8}$ & $0.49\pm0.19$ \\
4.0 & \phn1 & $(3.9\pm3.9)\times10^{-9}$ & \phd\phd$0.83\pm$\nodata\phn \\
\enddata
\end{deluxetable}
OH LFs previously computed by Baan (1991) and Briggs (1998) show similar
properties to the Arecibo OHM survey LF, as indicated in Figure \ref{hists3}.
Baan's OH LF samples a
larger range of \LOH, showing a knee at roughly $10^{2.5} L_\odot$.  The 
number of OHMs in the Arecibo sample contributing to the OH LF below 
$L_{OH}=10^{2.2} L_\odot$ is inadequate to confirm this turnover in the 
LF\@.  The survey sensitivity cutoff at these low line luminosities is 
severe, as seen in Figures \ref{survey_z} and \ref{hists3}.  Also noteworthy
is the higher OHM density found in the Arecibo LF versus Baan's 1991
result for $L_{OH}>10^3 L_\odot$.  
Briggs (1998) derived an OH LF from a quadratic OH-FIR relation combined 
with a 60$\mu$m luminosity function derived by Koranyi \& Strauss (1997).
Briggs computes the OH LF analytically assuming an OHM fraction of unity
in LIRGs for all $L_{60\mu m}$ (Figure \ref{hists3}b, {\it dotted line})
and numerically for an increasing OHM fraction versus $L_{60\mu m}$
(Figure \ref{hists3}b, {\it histogram}).
Although a quadratic OH-FIR relation is not supported by the known OHMs 
(Paper III), Briggs' OH LF follows the Arecibo OH LF data points remarkably
closely above $L_{OH}=10^{2} L_\odot$.  Briggs obtains a rough 
power-law slope of $-1.15$, which is inconsistent with the Arecibo result of
$-0.64\pm0.2$ (Equation \ref{eqn:LF}).  
The inconsistency can be 
attributed to the steep slope at the high luminosity end of the 60$\mu$m LF of 
\citet{kor97} compared to the shallower slope obtained by \citet{yun01}
and \citet{kim98}.  When translated into an OH LF, the steep 60$\mu$m slope 
is significantly lessened by Briggs's use of a quadratic OH-FIR relation.
The Arecibo power-law slope for OHMs 
{\it is} consistent with the power-law LF Kim \& Sanders (1998) 
determined for ULIRGs.  They found $\Phi \propto L_{IR}^{-2.35\pm0.3}$
Mpc$^{-3}$ mag\minusone\ over $L_{IR} = 10^{12}$--10$^{13} L_\odot$
(the exponent becomes $-0.94\pm0.12$ when $\Phi$ is expressed in Mpc$^{-3}$
dex\minusone).  

A well-determined OH luminosity function forms the foundation of any 
galaxy evolution study which uses OH megamasers as luminous radio tracers
of mergers, dust-enshrouded star formation, or supermassive black hole
binaries.  We now use this new OH LF to predict the detectability and 
abundance of OHMs available to deep radio surveys.

\section{Detecting OH Megamasers at High Redshift\label{sec:detect_hiz}}

OH megamasers are excellent luminous tracers of merging galaxies.  They
may become a tool for measuring the merging history of galaxies across
much of cosmic time and determine the contribution of merging supermassive
black holes to the low frequency gravitational wave background and to 
low frequency 
gravitational wave bursts.  They also provide an extinction-free tracer of
bursts of highly obscured star formation and may provide an independent 
measure of the star formation history of the universe.  Application of 
OH megamasers to these topics requires (1) that they be
detectable at moderate to high redshifts, and (2) that their sky surface
density be high enough for radio telescopes to detect at least a few
per pointing.  

\subsection{Detectability with Current and Planned Facilities}

\begin{figure}[!hb]
\epsscale{1}
\plotone{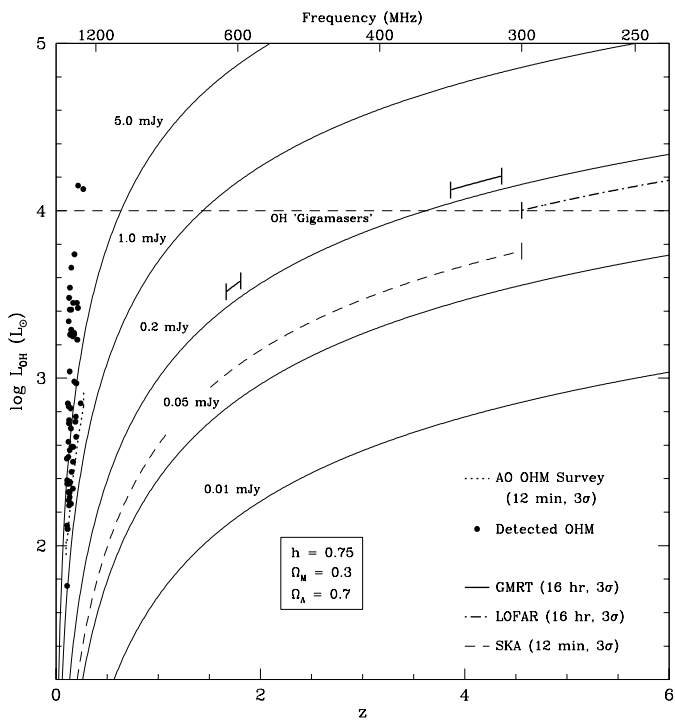}
\caption[Detectability of OH Megamasers]
{Detectability of OH megamasers.  Contours show the sensitivity required to 
detect an OHM with luminosity \LOH\ at redshift $z$.  Included are the 
results and sensitivity of the Arecibo OHM survey and predictions for the GMRT,
the SKA, and LOFAR.
\label{sensitivity}}
\end{figure}

The detectability any given OH megamaser at high redshift depends on 
the strength of the OHM, the sensitivity of the instrumentation, and 
cosmology.
Moving beyond $z\simeq0.2$ requires a careful treatment of cosmology
because the differences between luminosity distances and volumes among the
manifold of possible cosmologies become significant.
We assume for 
this analysis that 
$H_\circ = 75$ km s\minusone\ Mpc\minusone, $\Omega_M = 0.3$, and
$\Omega_\Lambda = 0.7$.  

Assumptions are also required to translate the observed quantity, flux 
density, to a line luminosity.  We assume that the integrated flux density
can be approximated by the product of the peak flux density and an average
rest-frame width, narrowed by the redshift:
\begin{equation}
F_{OH} = f_{OH}\, {\Delta\nu_\circ\over1+z} 
	= f_{OH}\,{\nu_\circ\,\Delta\mbox{v}_\circ\over c\,(1+z)}
\end{equation}
where the assumed rest-frame width is $\Delta\mbox{v}_\circ = 150$ 
km s\minusone.  

Figure \ref{sensitivity} plots sensitivity thresholds for a 3$\sigma$ 
detection of an OHM with luminosity \LOH\ at redshift $z$.  Included
are the Arecibo OHM survey detections and sensitivity limit, predictions
for a 16 hour integration by the Giant Metrewave Radio Telescope (GMRT) 
at 610 and 327 MHz, predictions for a 12 minute integration by the 
Square Kilometer Array (SKA) down to 300 MHz, and predictions for a 
16 hour integration by the Low Frequency Array (LOFAR) below 300 MHz.  
The GMRT is assumed to span 32 MHz with
0.12 MHz channels and obtain 0.083 mJy RMS noise per beam 
($T_{sys} \simeq 100$ K; Gain $= 0.32$ K 
Jy\minusone).\footnote{See 
http://www.ncra.tifr.res.in/ncra\_hpage/gmrt/gmrt\_spec.html.}
The SKA sensitivity 
is computed by scaling up the Arecibo OHM survey by a factor of 
25 in collecting area but keeping all other parameters the same.  Such 
a survey would have an RMS noise level of 0.027 mJy in 12 minutes
of integration per 49 kHz channel.  
The sensitivity of LOFAR below 300 MHz is estimated from guidelines 
provided by M. P. van Haarlem\footnote{See 
http://www.lofar.org/science/index.html.} 
to be 0.047 mJy RMS per 0.125 MHz channel in 16 hours of integration.  

Note that, RFI environment and available
receivers aside, Arecibo could detect OH gigamasers out to roughly $z=1$ 
in a 12 minute integration.  The GMRT 
shows much promise for the detection
of $L_{OH}>10^{3.5} L_\odot$ OHMs at $z=1.7$ in an integration of 
16 hours.  In integration times of less than an hour, a SKA would be 
able to detect a large fraction of the OHM 
population out to medium redshifts and all OH gigamasers down to its
lowest proposed operating frequency near 300 MHz.  
LOFAR would be able to detect OH gigamasers
in roughly 32 hours of integration from $z \simeq 4.5$ back to the 
reionization epoch if they exist.  
Clearly OH megamasers are detectable at moderate
redshifts with current facilities and at high redshifts with future
arrays, but how abundant might they be?

\subsection{The Sky Density of OH Megamasers}

The OH megamaser luminosity function (OH LF) derived in \S 
\ref{sec:OHLF} can predict the sky density of detectable OHMs 
as a function of instrument sensitivity, bandpass, and redshift.
A useful function in terms of observational parameters would
be the number of OHMs detected per square degree on the sky per
MHz bandpass searched.  This can be expressed as an integral of the 
OH LF over the range of detectable \LOH:
\begin{eqnarray}
{dN\over d\Omega\,d\nu} &=& \int_{\log L_{OH,min}(z)}^{\log L_{OH,max}}
		{dN\over d\Omega\,d\nu\,d\,\log L_{OH}}\,d\,\log L_{OH} 
								\nonumber\\
	&=& \int_{\dots}^{\dots} {dN\over dV\,d\,\log L_{OH}}\,
		{dV\over d\Omega\,d\nu}\,d\,\log L_{OH}
								\nonumber\\
	&=& {dV\over d\Omega\,d\nu}
		\int^{\dots}_{\dots}\Phi(L_{OH})\,d\,\log L_{OH}.
	\label{eqn:dNdOdnu}
\end{eqnarray}
Recall that the OH LF is the number of OHMs with luminosity \LOH\ 
per Mpc$^3$ per logarithmic interval in \LOH, expressed as 
\begin{equation}
	\Phi(L_{OH}) = b\,L_{OH}^a.
\end{equation}
Hence, the integral in Equation \ref{eqn:dNdOdnu} reduces to a simple
form
\begin{equation}
  \int^{\dots}_{\dots} \Phi(L_{OH})\,d\log L_{OH} 
	= {b\over a\ln 10} \left[L_{OH,max}^a - L_{OH,min}^a(z)\right].
\end{equation}
The volume element 
per unit solid angle per unit frequency can be translated into the 
usual panoply of cosmological parameters (Weinberg 1972):
\begin{eqnarray}
{dV\over d\Omega\,d\nu} &=& {dz\over d\nu}\,{dV\over d\Omega\,dz} \nonumber\\
	&=& {dV\over d\Omega\,dz}\,{(1+z)^2\over\nu_\circ} \nonumber\\
	&=& {c\,D_L^2\over H_\circ\nu_\circ
	\sqrt{(1+z)^3\Omega_M+\Omega_\Lambda}}\ .
\end{eqnarray}
Hence, the final result for the sky density of OHMs in a flat spacetime
has the form
\begin{eqnarray}
{dN\over d\Omega\,d\nu} &=& {c\,D_L^2\over H_\circ\nu_\circ
  \sqrt{(1+z)^3\Omega_M+\Omega_\Lambda}}\,{b\over a\ln 10} \nonumber \\
  & & \mbox{} \times \left[L_{OH,max}^a - L_{OH,min}^a(z)\right].
\end{eqnarray}
The minimum detectable \LOH\ depends on the sensitivity of observations and
the redshift, as shown in Figure \ref{sensitivity}.  The maximum \LOH\ 
is assumed to be $10^{4.4} L_\odot$
which is a factor of two larger than the luminosities of the known OH 
gigamasers.  The effects
of relaxing this conservative assumption are discussed below.  
When $L_{OH,min} = L_{OH,max}$ there are no OHMs left to detect.

\begin{figure}[ht]
\epsscale{1}
\plotone{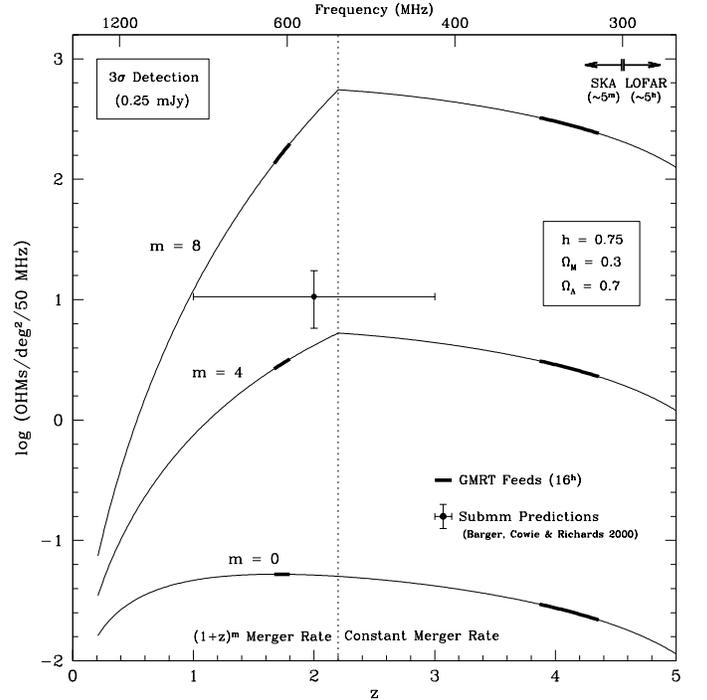}
\caption[Sky Density of Detectable OH Megamasers]
{Sky density of detectable OH megamasers at 0.25 mJy.
The expected detections for several galaxy merger histories 
are labeled by the rate of increase in the merger rate as 
$(1+z)^m$.  The point with error bars is the prediction for 
OHM detections based on the space density of submillimeter detections 
computed by Barger, Cowie, \& Richards (2000).  
The turnover in merger rate from increasing to constant at $z=2.2$ is
indicated by the dotted vertical line.
The GMRT (currently accessible bands are indicated in bold) can
reach the RMS noise of 0.083 mJy in each 0.12 MHz channel in roughly 16 
hours of integration.  The proposed SKA might reach the 
same noise level in a few minutes of integration while LOFAR would require 
several hours.  
\label{detect_general}}
\end{figure}

Finally, we fold in an evolution parameter $m$ which scales the sky 
density of OHMs as $(1+z)^m$.  Assuming $H_\circ = 75$ km s\minusone\ 
Mpc\minusone, $\Omega_M = 0.3$, and $\Omega_\Lambda = 0.7$, 
and folding in the OH LF 
parameters $a$ and $b$ from \S \ref{sec:OHLF}, we obtain the sky
density of OHMs for any sensitivity level as a function of redshift.
The detectable sky density of OHMs versus redshift is plotted in Figure
\ref{detect_general} for a 3$\sigma$ OH line detection at 0.25 mJy.  
Several evolution scenarios are plotted:  $m=0$ (no evolution), $m=4$, 
and $m=8$.  We assume an arbitrary turnover in the evolution factor
at $z=2.2$, after which the number density of OHMs is constant.  This
turnover in the merger rate corresponds roughly to the turnover in 
the QSO luminosity function.  
The point with error bars is a prediction of the sky density of OHMs 
derived from the space density of submillimeter detections 
(which are ostensibly ULIRGs) computed by Barger, Cowie, \& Richards (2000)
assuming that half of the submillimeter galaxies produce detectable
OHMs.
Bands currently accessible to the GMRT (610 and 327 MHz) 
in roughly 16 hours of integration
are indicated in bold.  The proposed Square Kilometer Array might reach 
the same sensitivity level in a few minutes of integration.
If the submillimeter predictions are to 
be believed, very strong evolution is favored and the GMRT could detect 
dozens of OHMs per square degree in its 32 MHz bandpass in a reasonable
integration time.  
This is a very exciting prospect and can indicate 
which models of galaxy evolution are favored, as discussed below.

Briggs (1998) makes predictions of the sky density of OHMs using the 
FIR LF of LIRGs of Koranyi \& Strauss (1997), a quadratic OH-FIR relation, 
and an optimistic OHM fraction as a function of $L_{60\mu m}$ derived from 
Baan (1989) and Baan \etal (1992).  Briggs predicts roughly 16 OHMs 
at $z=2$ per
square degree per 50 MHz bandpass for $m=4.5$ and a sensitivity level of 0.2
mJy.  The predictions of OHM detections based on the Arecibo survey OH LF
are not as rosy; only about 6 OHMs would be detected for $m=4.5$ at 0.2 mJy
at $z=2$.
The main discrepancy between the two predictions lies in Briggs' use
of a quadratic OH-FIR relation and a high OHM fraction in LIRGs compared
to the fraction detected in the Arecibo OHM survey.  

If the assumptions built into these detection predictions are changed, the 
results change only slightly below $z\sim2$.  
For example, if the OH LF is extended to 
$L_{OH}=10^5 L_\odot$, there is no significant effect on the detection
rates below $z\sim3$ because the most luminous OHMs are rare.  The detection
rates at the highest redshifts will change significantly because only the 
tail of the OH LF is detected, but predictions beyond $z=2$ are wild 
speculations with no supporting data.  The choice of cosmology also does not
alter the detection rates dramatically.  Changing from an 
$\Omega_\Lambda =0$ to an $\Omega_\Lambda =0.7$ universe amounts to roughly
a factor of 1.5 increase in OHM detections at $z<2$.  The effect again 
becomes more significant at higher redshifts.



\subsection{The Merging History of Galaxies}

A property as basic as the number density of galaxies as a function of
redshift is not well known.  The number density of galaxies depends
of course on the merging history of galaxies, which is an essential
ingredient in theories of structure formation and galaxy evolution.  
There have been a number of studies which attempt to measure the
merger rate or merger fraction of galaxies versus redshift, and most
of them parameterize the evolution of mergers with a factor $(1+z)^m$.
A {\it Hubble Space Telescope} survey identifying optical mergers 
and close pairs of galaxies 
indicates $m=3.4\pm0.6$ from $z=0$ to $z=1$ (Le F\'{e}vre  \etal 2000), 
whereas the 1 Jy ULIRG survey indicates an exponent of $m=7.6\pm3.2$ 
(Kim \& Sanders 1998) which is similar to the evolution of bright QSOs 
with redshift as $m\sim6$ up to $z\sim2.5$
(Briggs 1998; Hewett \etal 1993; Schmidt \etal 1995).  Semi-analytic
models of hierarchical galaxy formation by Kauffmann \& Haehnelt (2000)
indicate an evolution of the number density of gas-rich mergers from 
$z=0$ to $z=2$ roughly following $m=4$.  The sampling of work on this
topic listed here is by no means complete.  In general, most studies
indicate a strong increase of the merging rate with redshift, but it
is likely to be a strong function of the total dark halo mass 
(for which bolometric luminosity is a proxy) and the mass ratio of the 
merging pair \citep{kho01}.
A luminosity dependence is seen in the evolution of quasars and LIRGs, 
with the most luminous objects showing the strongest evolution 
(Schmidt \etal 1995; Kim \& Sanders 1998).  

Although a deep survey for OH megamasers may not be translatable into
a merger {\it rate}, it does discriminate between various galaxy
evolution scenarios.  For example, there are two orders of magnitude 
difference between the OHM detections expected from $m=8$ versus $m=4$
at $z\sim2$.  Even a lack of detections in an adequately deep OHM 
survey would provide new constraints on the evolution of mergers with
redshift and would suggest a revision of the conventional interpretation
of submm galaxies as high-redshift ULIRGs.  If, on the other hand, OHMs
were detected in a deep survey, this would not only indicate the 
evolution of merging but also provide an extinction-free redshift 
determination for submillimeter galaxies which have proven to be extremely 
difficult to identify and observe at optical wavelengths or even IR wavelengths
(\eg --- Smail \etal 2000; Ivison \etal 2000).

\subsection{The Star Formation History of the Universe}

Merging systems traced by OH megamasers are in the throes of extreme
starbursts, but much of the star formation is highly obscured by dust.  
The history of obscured star formation is difficult to determine with 
current instruments which detect the reprocessed light directly 
at FIR or submm wavelengths due to poor sensitivity or the
difficulty of determining redshifts, respectively.  OH megamasers
are a promising solution to these limitations because they
are luminous, are unattenuated by dust, and provide redshifts.
As Townsend \etal (2001) point out, OH megamasers may be used
to determine the nature and evolution of submm galaxies and to indicate
their relevance to the star formation history of the universe.  
If the connections between OHMs and their hosts, particularly between
\LOH\ and \LFIR\ (or at least the OH LF and the FIR LF), are 
valid at moderate to high redshifts, then surveys for OHMs can be used
as an independent measure of the star formation history 
across a large span of the age of the universe.

\subsection{Constraining The Gravitational Wave Background}\label{subsec:gwb}

One final potential application of OHMs addresses the end result of 
massive mergers:  the formation and coalescence of binary supermassive
black holes.  Supposing that each galaxy in a merger contains a supermassive
black hole, the rapid coalescence of nuclei due to dynamical friction
will produce a binary supermassive black hole which will continue to 
decay until a final coalescence event which will produce (among other things)
a burst of gravitational waves.  Bursts from supermassive black holes 
are likely to be the major source of $10^{-5}$--$10^0$ Hz gravitational waves
(Haehnelt 1994) which may someday be detectable by the Laser Interferometer 
Space Antenna (LISA) or long-duration pulsar timing.  
The merging rate of 
galaxies, as well as the masses of the black holes involved, determine
the event rate detectable by LISA.  The integrated merging history of 
galaxies determine the noise levels produced in pulsar timing, and would
provide a ``foreground'' to any cosmological background of gravitational
waves produced during the inflationary epoch (D. Backer 2001, private 
communication).  Clearly,
getting some handle of the event rate of supermassive black hole mergers
would provide much needed constraints and thresholds for the difficult
work of detecting gravitational waves.

\section{Summary}

The OH luminosity function constructed from a sample of 50 OH megamasers
detected by the Arecibo OH megamaser survey
indicates a power-law falloff with increasing OH luminosity
\begin{displaymath}
  \Phi = (9.8^{+31.9}_{-7.5}\times10^{-6})\ L_{OH}^{-0.64\pm0.21}\ \
		\mbox{Mpc$^{-3}$ dex$^{-1}$}
\end{displaymath}
valid for $2.2<\log L_{OH}<3.8$ (expressed in $L_\odot$) and $0.1<z<0.23$.
The OH LF is used to predict the areal density of detectable OHMs at 
arbitrary redshift for a manifold of galaxy merger evolution scenarios
parameterized by $(1+z)^m$ where $0 < m < 8$.  For reasonable 
choices of $m$, an ``OH Deep Field'' 
obtained with the Giant Metrewave Radio Telescope at 610 MHz ($z=1.73$)
may detect dozens of OHMs per square degree in a reasonable integration
time.  A lack of detections in a sufficiently deep field would also 
significantly constrain the evolution of merging and exclude the most
extreme evolution scenarios.

\acknowledgements

The authors are very grateful to Will Saunders for access to the PSCz catalog
and to the excellent staff of NAIC for observing assistance and support.  
We thank the anonymous referee for thoughtful comments and suggestions.
This research was supported by Space Science Institute archival grant 
8373 and NSF grant AST 00-98526
and made use of the NASA/IPAC Extragalactic Database (NED) 
which is operated by the Jet Propulsion Laboratory, California
Institute of Technology, under contract with the National Aeronautics 
and Space Administration.

\end{document}